\documentclass[aps,prl,reprint,superscriptaddress,tightenlines,amsmath,amssymb]{revtex4-2}
\usepackage{amsmath,amssymb,amsfonts,amsthm}
\usepackage{graphicx}
\usepackage[T1]{fontenc}
\usepackage[colorlinks=true,citecolor=blue,linkcolor=blue,urlcolor=blue]{hyperref}
\usepackage{braket}
\usepackage{xcolor}

\usepackage[english]{babel}
\usepackage[autostyle, english = american]{csquotes}
\MakeOuterQuote{"}

\begin{document}

\title{Typical Entanglement of Superpositions}

\author{Damien Quinn}
\email{damien.quinn.2026@mumail.ie}
\affiliation{Department of Physics, Science Building, Maynooth University, Maynooth, Co. Kildare W23 F2H6, Ireland}

\author{Joshuah T. Heath}
\email{joshuah.heath@mu.ie}
\affiliation{Department of Physics, Science Building, Maynooth University, Maynooth, Co. Kildare W23 F2H6, Ireland}
\affiliation{Hamilton Institute, Eolas Building, Maynooth University, Maynooth, Co. Kildare W23 A3HY, Ireland}

\author{Graham Kells}
\email{graham.kells@mu.ie}
\affiliation{Department of Physics, Science Building, Maynooth University, Maynooth, Co. Kildare W23 F2H6, Ireland}
\affiliation{Hamilton Institute, Eolas Building, Maynooth University, Maynooth, Co. Kildare W23 A3HY, Ireland}
\affiliation{Dublin Institute for Advanced Studies, School of Theoretical Physics, 10 Burlington Road, Dublin, Co. Dublin D04 C932, Ireland}

\date{\today}

\begin{abstract}
\noindent We investigate universal entanglement properties inherent to superpositions of randomized states. We find that an $m$-fold superposition of typical states may be classified into two distinct entanglement classes via the 2nd R\'enyi entropy density $s_2$. The maximally entangled regime is defined by $s_2 \sim \ln (2)$, for which superposition adds no additional entanglement. The sub-maximally entangled regime, $s_2<\ln 2$, instead constrains the reduced density matrices of independent components to be orthogonal in the thermodynamic limit, which fixes the entanglement of the superposition to a logarithmic enhancement $\Delta S(m)=\ln (m)$. As a consequence, an exponentially large number of superpositions is required to transition from the sub-maximally entangled class to maximal entanglement. We explicitly calculate $s_2$ and the logarithmic enhancement, and demonstrate orthogonality for two canonical examples of the sub-maximally entangled regime (superpositions of pure Gaussian states and of random matrix-product states). We also examine the entanglement of superpositions of random stabilizer states, and discuss their relaxation to the Haar limit.
\end{abstract}

\maketitle

\textit{Introduction}-- A pure state drawn at random from a many-body Hilbert space is, with overwhelming probability, almost maximally entangled across any bipartition~\cite{Page1993Aug,FoongKanno1994,SanchezRuiz1995,Sen1996}. This typicality~\cite{Goldstein2006Feb,Popescu2006Nov,Reimann2007Oct,Wilde2017Feb} result is the reference point for the entanglement of more structured random ensembles; e.g., random Gaussian states~\cite{Magan2016Jan,Bhattacharjee2021Dec,bhk,Bianchi2022Jul,Yu2023Jan, Swietek2026Jul}, random matrix product states (MPS) ~\cite{garnerone2010a,garnerone2010b,haferkamp2021}, and random stabilizer states~\cite{SmithPRA2006}. Whereas the typical entanglement of random stabilizer states saturates within $\mathcal{O}(1)$ of the Page value in the thermodynamic limit, the typical entanglement of Gaussian states and finite-bond-dimension random MPS remains significantly below maximal entanglement regardless of system size.
In this latter sub-maximal class, an interesting question is if the additional entanglement of $m$ superpositions $\ket{\Psi_m} = \sum_{i=1}^m \lambda_i\ket{\psi_i}$ may enhance the total entanglement~\cite{LindenPRL2006,GourPRA2007,Gour2008Jan} to the maximally allowed value. Superpositions of such states are ubiquitous in many-body physics, and feature prominently in quantum chemistry~\cite{Szabo1996Jul,McArdle2020Mar,EIDOS}, quantum information processing~\cite{Cudby2024,ReardonSmith2024,DiasQuantum2024}, and condensed matter physics~\cite{Anderson1973Feb,Baskaran1988Jan,Misguich2005Jan,BoutinPRR2021}.

We show that superpositions of typical states fall into two broad classes, set by the 2nd R\'enyi entropy density $s_2$ of the components. In the \emph{sub-maximally entangled} class, $s_2<\ln (2)$, the reduced density matrices (RDMs) of independent components are forced to be \emph{one-sided orthogonal}~\cite{GourPRA2007,Gour2008Jan} in the thermodynamic limit; i.e., the cross-overlap of two component RDMs is exponentially smaller than their self-overlap, at the rate $c_\perp\equiv\ln (2)-s_2>0$. This orthogonality fixes the entanglement of an $m$-component superposition to a universal logarithmic enhancement of the 2nd R\'enyi entanglement $\langle S(m)\rangle$,
\begin{equation}
\Delta S_2(m)\equiv\langle S_2(m)\rangle-\langle S_2(1)\rangle  = \ln(m),
\label{eq:main_result}
\end{equation}
a universal upper bound for any system in the sub-maximal class. The $\ln (m)$ scaling is a manifestation of the Linden--Popescu--Smolin (LPS) bound~\cite{LindenPRL2006,GourPRA2007,Gour2008Jan}.

\begin{figure}[t]
\includegraphics[width=\columnwidth]{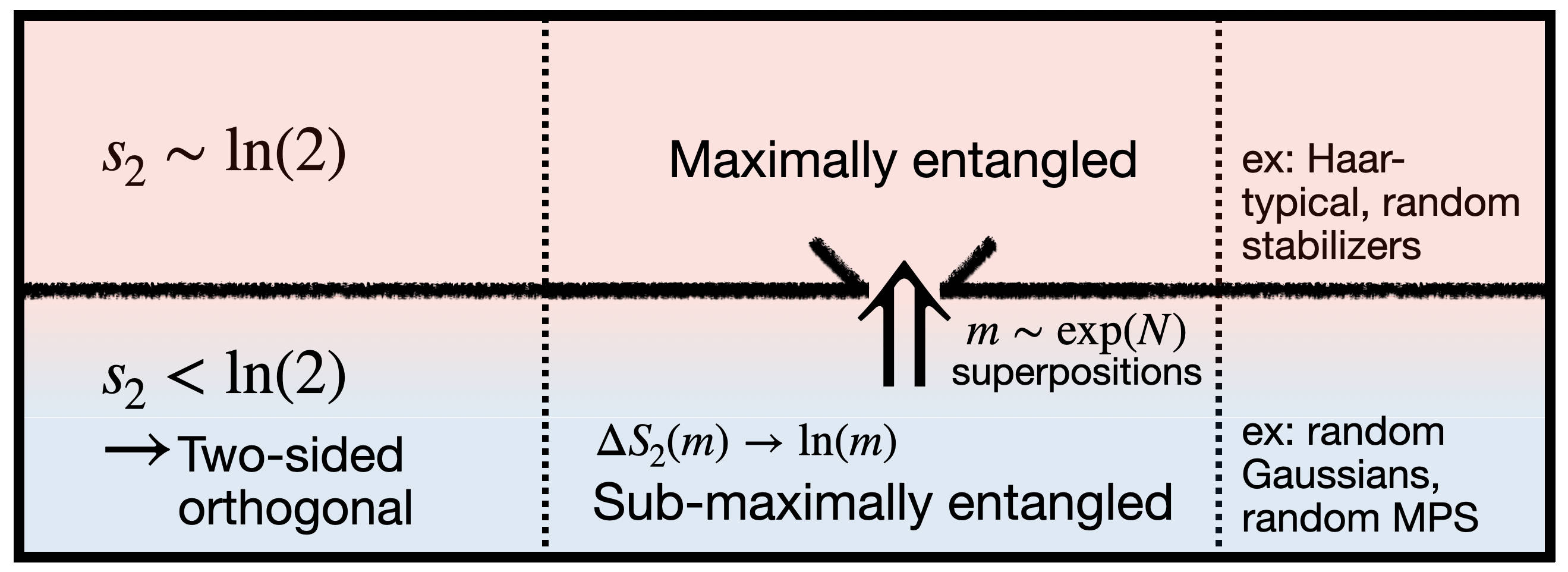}
\caption{Schematic of the two entanglement classes for typical superpositions. On the bottom is the sub-maximally entangled class, defined by a 2nd Rényi entanglement entropy $s_2<\ln(2)$. These systems are shown to be {\it constrained} to be two-sided orthogonal~\cite{LindenPRL2006}. Canonical examples of this entanglement class include random Gaussian states and random matrix-product states (MPS). On the top is the maximally entangled class, defined by $s_2\sim\ln(2)$. These systems are not two-sided orthogonal, and canonical examples include Haar-typical and random stabilizer states. An exponential number of superpositions $m\sim \exp(N)$ is required to "break free" from the sub-maximally entangled class to the maximal, where $N$ is the system size.
}
\label{fig:firstpage}
\end{figure}

\begin{figure}[t]
\includegraphics[width=\columnwidth]{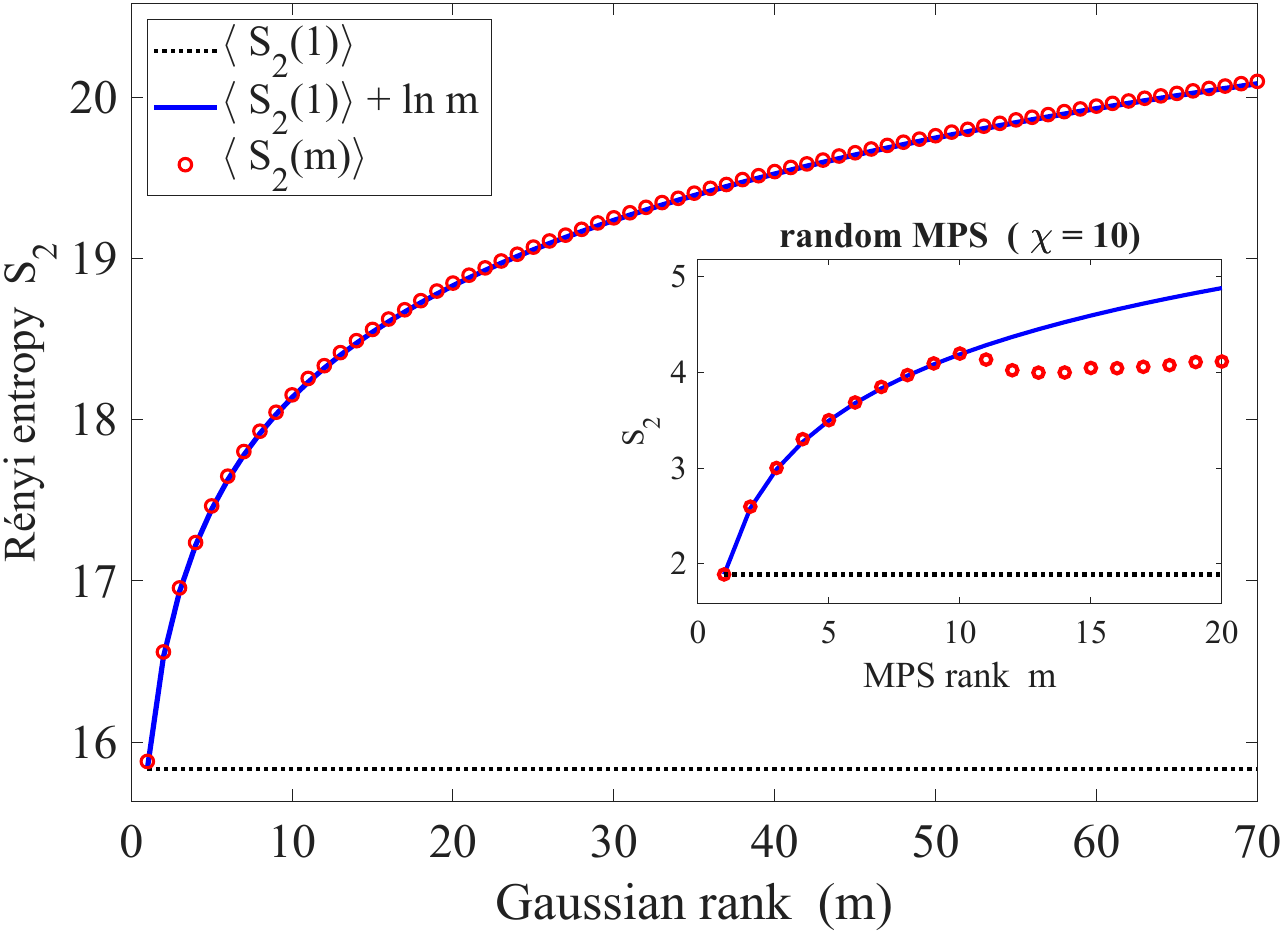}
\caption{The 2nd Rényi entanglement entropy $\langle S_2(m)\rangle$ versus number of Gaussian components~$m$ for a system size of $N=100$ at equal bipartition. The dashed baseline is the single-component average $\langle S_2(1)\rangle$. The $\ln(m)$ growth predicted by Eq.~\eqref{eq:main_result} is confirmed up to $m = 70$. (Inset) $S_2$ entanglement entropy of superpositions of $m$ random MPS ($d = 2$, $N = 100$, $\chi = 10$) with total bond-dimension cutoff $\chi_{\max} = 100$.  The $\ln(m)$ scaling holds while $m\chi \leq \chi_{\max}$. Beyond this regime, the entropy plateaus at $S_2 \approx \ln(d\chi_{\max}/(d{+}1))$. }
\label{fig:scaling}
\end{figure}

The two entanglement classes are populated by well-known ensembles. In the sub-maximal class, random fermionic Gaussian states have entanglement density $s_2 =  - \ln\!\left((3+2\sqrt{2})/8\right)$ at equal bipartition (from a Wachter law~\cite{wachter,Wachter1980Sep}), and random matrix-product states of bond dimension $\chi$ and physical dimension $d$ have $S_2=\ln(d\chi/(d{+}1))$ (from a Marchenko--Pastur law~\cite{MarchenkoPastur1967}). Both are well below the Haar value $\sim \ln(2)$~\cite{Page1993Aug}, and thus gain $\ln (m)$ entanglement under superposition. In the maximal class, random stabilizers sit at $s_2 \sim \ln (2)$, and superposition contributes intensive $\mathcal{O}(1)$ entanglement. Under superposition, deviation from the Haar value is closed via parameter-free, $N$-independent scaling laws.

\textit{Orthogonality and entanglement scaling for typical superpositions.---}The $\ln (m)$ scaling law rests on a logical chain of equivalent conditions on the components of the ensemble; namely, a sub-maximal entanglement density (i.e., $s_2<\ln (2)$) forces the component RDMs to be one-sided orthogonal (at rate $c_\perp=\ln (2)-s_2$),  which in turn fixes $\Delta S_s(m)=\ln (m)$. We establish this chain of logic below for a generic random ensemble. We then return to the canonical examples of random fermionic Gaussian states and MPS to analytically derive $s_2$ in each case. 

Consider a superposition $\ket{\Psi_m}=\sum_{i=1}^m\lambda_i\ket{\psi_i}$ on a bipartition $AB$, where the components are drawn independently from a fixed random ensemble. We measure entanglement by the R\'enyi entropies of the reduced state $\rho_A=\mathrm{Tr}_B\ket{\Psi_m}\!\bra{\Psi_m}$,
\begin{equation}
S_\alpha=\frac{1}{1-\alpha}\ln\mathrm{Tr}(\rho_A^\alpha),\qquad S_2=-\ln\mathrm{Tr}(\rho_A^2),
\label{eq:renyi_def}
\end{equation}
working throughout at second order, with density $s_2\equiv S_2/|A|$.  The purity itself is a quadratic form in the components,
\begin{equation}
\begin{aligned}
\mathrm{Tr}(\rho_A^2)&=\sum_{ijk\ell}\lambda_i\lambda_j^*\lambda_k\lambda_\ell^*\,\mathcal{T}_{ijk\ell},\\
\mathcal{T}_{ijk\ell}&=\mathrm{Tr}\!\big[\mathrm{Tr}_B(\ket{\psi_i}\bra{\psi_j})\,\mathrm{Tr}_B(\ket{\psi_k}\!\bra{\psi_\ell})\big],
\end{aligned}
\label{eq:purity_tensor}
\end{equation}
where $\mathcal{T}_{ijk\ell}$ is the \emph{purity tensor}. Our central observation is that whenever the components of $\mathcal{T}_{ijk\ell}$ are sub-maximally entangled, the purity tensor becomes diagonal in the thermodyanamic limit, i.e. $\mathcal{T}_{ijk\ell}\to\delta_{ij}\delta_{k\ell}\delta_{ik}\,e^{-S_2^{(i)}}$, so that for equal weights $\lambda_i=1/\sqrt m$,
\begin{equation}
\mathrm{Tr}(\rho_A^2)\approx\frac1{m^2}\sum_i e^{-S_2^{(i)}}=\frac1m\,\langle e^{-S_2}\rangle\approx\frac1m\,e^{-\langle S_2\rangle},
\label{eq:diag_purity}
\end{equation}
and the entanglement therefore grows by $\ln (m)$ [Eq.~\eqref{eq:main_result}]. Such {\it diagonal dominance} of the purity tensor is determined by the difference of the component entropy density $s_2$ and the Haar value $\ln(2)$. We will now elaborate further on the underlying physics which results in the diagonal dominance of $\mathcal{T}_{ijk\ell}$.

The diagonal and off-diagonal elements of the purity tensor are the \emph{self-} and \emph{cross-overlaps} of the component RDMs. By Schur's lemma~\cite{Georgi,Serre,Etingof2009Jan}, the ensemble-averaged RDM is maximally mixed, $\overline{\rho^A}=I/2^{|A|}$, so two independent RDM components of the ensemble have cross-overlap $\mathbb{E}[\mathrm{Tr}(\rho_\psi^A\rho_\phi^A)]=2^{-|A|}$, fixed by the Hilbert space dimension alone. However the self-overlap $\mathbb{E}[\mathrm{Tr}(\rho_\psi^A)^2]=e^{-s_2|A|}$ measures the component entanglement. Their ratio,
\begin{equation}
\frac{\mathbb{E}[\mathrm{Tr}(\rho_\psi^A\rho_\phi^A)]}{\mathbb{E}[\mathrm{Tr}(\rho_\psi^A)^2]}
\sim C_0\,e^{-c_\perp|A|+o(|A|)}\xrightarrow{|A|\to\infty}0,
%\qquad(s_2<\ln 2),
\label{eq:orthogonality}
\end{equation}
vanishes whenever $c_\perp=\ln 2-s_2>0$, with $C_0=\mathcal{O}(1)$ accounting for any symmetry that would leave $\overline{\rho^A}$ maximally mixed only within sectors.

A vanishing ratio Eqn.~\eqref{eq:orthogonality} is equivalent to the one-sided orthogonality of the component RDMs on $A$~\cite{LindenPRL2006,GourPRA2007,Gour2008Jan}. The ratio weighs the cross-overlap of two distinct RDMs against the self-overlap of one RDM, so a vanishing ratio means that the cross-overlap is exponentially smaller than a component's own purity, leaving the RDMs orthogonal (see Appendix~\ref{app:criterion}). The same holds on $B$, so at an equal bipartition the generic ensemble states are \emph{two-sided orthogonal} \footnote{We use ``two-sided orthogonality'' rather than LPS's ``biorthogonality'' to avoid confusion with the non-Hermitian usage.}. Physically, the components of the RDM become perfectly distinguishable within a single subsystem; i.e., the two component reduced states overlap as $e^{-c_\perp|A|}$. As such, if RDM orthogonality holds, an observer with access to $A$ alone can tell the RDMs apart with exponentially improving precision as $|A|$ grows via local measurements.

Two-sided orthogonality is what renders the purity tensor diagonally dominant~\footnote{Two-sided orthogonality is also necessary for a universal closed form, as illustrated by the Bell states $\tfrac1{\sqrt2}(\ket{00}\pm\ket{11})$~\cite{LindenPRL2006}, each maximally entangled with $\mathrm{Tr}(\rho_\phi^A\rho_\psi^A)=\mathrm{Tr}(\rho_\phi^B\rho_\psi^B)=\tfrac12$, yet superposing to an unentangled state.}. The off-diagonal entries of $\mathcal{T}_{ijk\ell}$ are the cross-overlaps of Eqn.~\eqref{eq:orthogonality} which are shown to be negligible for large $|A|$: the $A$-side overlap $\mathrm{Tr}(\rho_A^{(i)}\rho_A^{(k)})$ on the \emph{direct} pairing ($i{=}j,\,k{=}\ell$) and the $B$-side overlap $\mathrm{Tr}(\rho_B^{(i)}\rho_B^{(k)})$ on the \emph{swap} pairing ($i{=}\ell,\,k{=}j$), each $2^{-|A|}$ against a diagonal self-purity $e^{-s_2|A|}$ (Appendix~\ref{app:patterns}). The $A$-side suppression gives the components orthogonal support, and the $B$-side leaves the reduced state an incoherent mixture (Appendix~\ref{app:two_faces}), so the purity tensor collapses onto its diagonal, Eq.~\eqref{eq:diag_purity}.

This collapse gives $\Delta S_2(m)=\ln (m)$ (an example of the LPS bound), which saturates as $|A|\to\infty$ and reaches $s_2=\ln (2)$ only with exponentially many superposed components. The von~Neumann result holds exactly, without the concentration step (Appendix~\ref{sec:entropy}). Unequal weights replace $\ln (m)$ by the collision entropy of the weights (Appendix~\ref{sec:nonequal}).

{\it Computing $s_2$: Fermionic Gaussian states --} The previous discussion allows us to convert the problem of entanglement scaling of typical superpositions into the computation of the entanglement entropy density $s_2$ for the component ensemble. We will first consider the entanglement scaling in a canonical example of the sub-maximally entangled class: random fermionic Gaussian states.

\begin{figure}[t]
\includegraphics[width=\columnwidth]{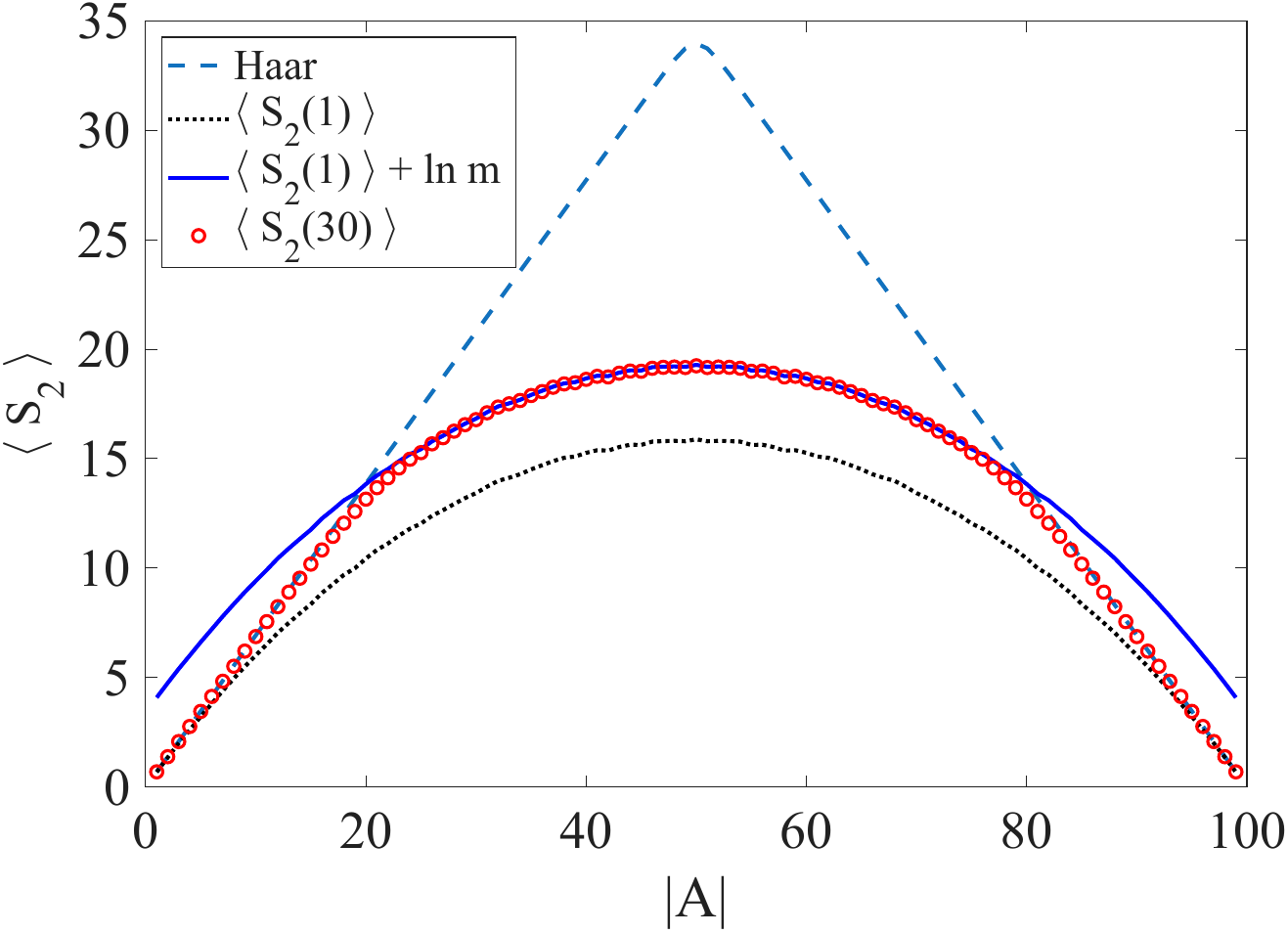}\vspace{5mm}\\
\includegraphics[width=\columnwidth]{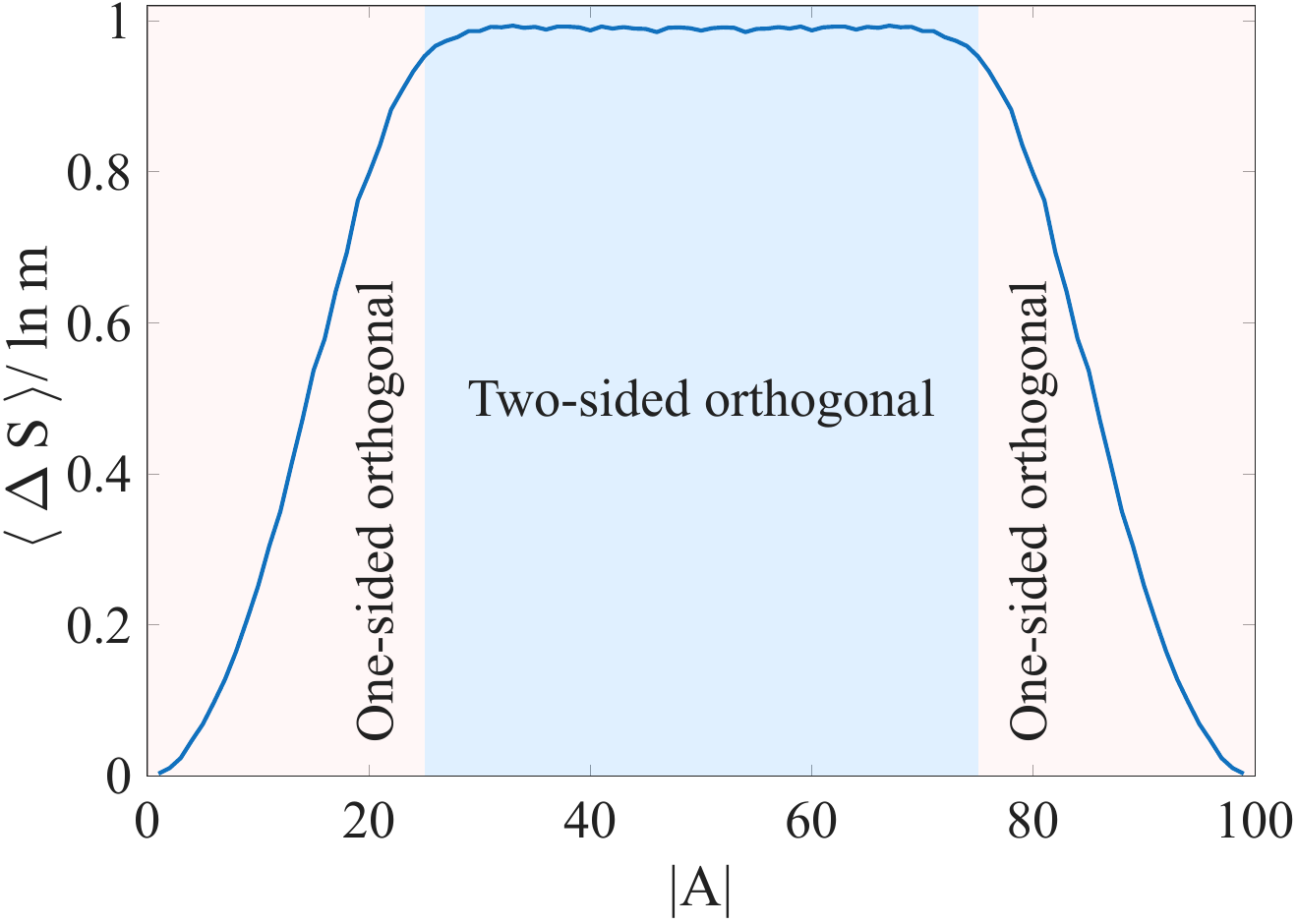}
\caption{Entanglement profile across all partitions for superpositions of Gaussian states ($N = 100$). (Top) $\langle S_2\rangle$ versus partition size for single components (black dotted) and $m=30$-component superpositions (red), compared against the Haar value (dashed blue). The excess $\langle S_2(1) \rangle+\ln(m)$ is also shown (solid blue).  (Bottom) The excess $(\langle S_2(m)\rangle  - \langle S_2(1) \rangle)/(\ln (m))$, saturating at unity for bulk partitions and falling off near the edges, where the smaller subsystem is the binding constraint for two-sided orthogonality. The excess plotted in this figure may therefore be interpreted as a metric for two-sided orthogonality for $m$-superpositions.}
\label{fig:cut_profile}
\end{figure}

 A fermionic Gaussian state on $N$ modes is characterized by a $2N\times 2N$ real antisymmetric covariance matrix $\Gamma$ with $\Gamma^2 = -I_{2N}$~\cite{BravyiPhysRevA2005}. We define a Haar-random Gaussian state to correspond to $\Gamma = O\Gamma_0 O^\top$ with Haar-distributed $O\in O(2N)$. Such random Gaussian states are constrained to be one-sided orthogonal, as expected for a member of the sub-maximally entangled class (see Appendix~\ref{sec:proof} for a rigorous proof). The universal result for the cross-overlap is a consequence of the irreducibility of the $\mathrm{Spin}(2N)$ Fock representation on the even-parity sector. Since number-conserving (i.e., Slater) ensembles are maximally mixed only within particle-number sectors, these ensembles exhibit an $\mathcal{O}(1)$ prefactor (given by the ratio defined in Eqn.~\ref{eq:orthogonality}) of $C_0\to 2/\sqrt{3}$ at half-filling (as opposed to $C_0=1$ for generic Bogoliubov de Gennes (BdG) states). From Peschel's factorization~\cite{peschel}, we may write the typical Rényi purity as $\mathrm{Tr}((\rho^A)^2) = 2^{-|A|}\prod_k(1+\epsilon_k^2)$, where $\pm i\epsilon_k$ are the eigenvalues of the covariance subblock $\Gamma^A$, so the entropy density is $s_2 = \ln 2 - |A|^{-1}\sum_k \ln(1+\epsilon_k^2)$. For Haar-random Gaussian states the $\{\epsilon_k^2\}$ form a Jacobi (MANOVA) ensemble~\cite{bhk,wachter} whose distribution is observed to converge to the Wachter law, and the linear statistic $|A|^{-1}\sum_k \ln(1+\epsilon_k^2)$ self-averages~\cite{johansson} to a strictly positive constant. Hence, $s_2<\ln 2$, and the components are one-sided orthogonal with exponential precision.  At equal bipartition, the Wachter law reduces to the arcsine law on $[0,1]$ for both BdG and number-conserving ensembles~\cite{bhk,bianchi}, and the integral evaluates to a closed form,
\begin{equation}
s_2 = \ln (2) - \ln\!\left(\frac{3+2\sqrt{2}}{4}\right)
\approx 0.317,
\label{eq:s2_gauss}
\end{equation}
This is $46\%$ of the maximal $\ln 2$, with von~Neumann density $\approx 0.386$. Equivalently, the orthogonality rate is $c_\perp = \ln (2) - s_2 = \ln\!\left((3+2\sqrt{2})/4\right) \approx 0.376$.  Because Haar-random Gaussian states have volume-law entanglement, $s_2$ depends only on the ratio $|A|/N$, not on any specific geometry or spatial dimension.  Fig.~\ref{fig:scaling} confirms the resulting $\ln(m)$ growth at $N = 100$ up to $m = 70$, with the purity tensor evaluated by the Pfaffian method of Boutin and Bauer~\cite{BoutinPRR2021} (see Appendix~\ref{sec:boutin_bauer}). For number-conserving Gaussian states, the same elements reduce to determinants and no Pfaffian machinery is needed.  Fig.~\ref{fig:cut_profile} resolves the gain across all parititions: the excess saturates at $\ln (m)$ for bulk partitions and falls off near the edges, where the smaller subsystem is the binding constraint. As such, there is a finite region about equal bipartition where two-sided orthogonality (and, as a consequence, logarithmic scaling of the entanglement entropy) persists for a large but finite-size system.

\begin{figure}[t]
\includegraphics[width=\columnwidth]{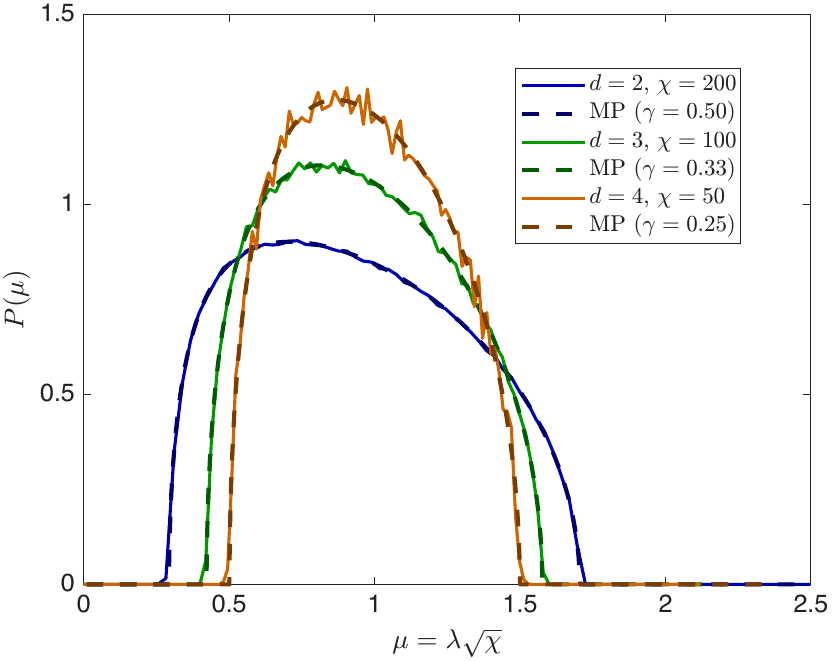}
\caption{Schmidt spectrum for a random MPS for a few example bond dimensions and aspect ratios. The spectrum collapses to a Marchenko--Pastur (MP) distribution for local dimensions $d = 2, 3, 4$ ($\gamma = 1/d$, $\sigma^2 = 1$; dashed curves).  As $d$ increases, the distribution concentrates, reflecting the fourth moment $\kappa = (d{+}1)/d \to 1$.}
\label{fig:mp_multid}
\end{figure}

The $\ln (m)$ scaling of the entanglement entropy for random Gaussian states has an immediate interpretation as a resource for simulation.  For equal-weight Gaussian superpositions, the fermionic Gaussian extent $\xi$ (i.e., a non-Gaussianity monotone that dictates classical simulation cost~\cite{DiasQuantum2024}) equals the number of components, $\xi = m$, so Eq.~\eqref{eq:main_result} gives
\begin{equation}
\langle S\rangle
= \langle S^{\mathrm{Gaussian}}\rangle + \ln(\xi),
\label{eq:extent}
\end{equation}
exactly for the von~Neumann entropy and to $\mathcal{O}(1/N)$ for the 2nd Rényi entanglement entropy.  Since the simulation algorithms of Refs.~\cite{Cudby2024,ReardonSmith2024,DiasQuantum2024} scale polynomially in $\xi$, Eq.~\eqref{eq:extent} is the entanglement counterpart of the extent; namely, states of polynomial extent carry only $\mathcal{O}(\ln (N))$ excess entanglement above the Gaussian baseline, and Haar-typical entanglement demands exponential $\xi$. We note here that recently Ref.~\cite{Swietek2026Jul} reached a similar conclusion for fermionic Gaussian superpositions. This static, extent-resolved statement is the entanglement counterpart to the dynamical picture of doped matchgate circuits~\cite{Terhal2002Mar,Jozsa2008Dec,Valiant2012Feb,Projansky2024Aug}, where ballistic and ultimately volume-law entanglement growth is restored only once an extensive number of non-Gaussian gates is injected~\cite{Paviglianiti2025Doped}. In this case, an extensive gate count is the circuit counterpart of exponential extent. 

In addition to the Gaussian extent, our main result for Gaussian states has implications in quantum chemistry. When applied to configuration-interaction (CI) expansions in quantum chemistry~\cite{Szabo1996Jul,EIDOS} (i.e., a state written as $m$ optimized Slater determinants), our logarithmic scaling argument predicts a maximum $\sim\ln (m)$ excess entanglement above the single-determinant baseline. As such, truncated CI cannot reach the maximal extensive entanglement of strongly correlated regimes without exponential $m$.

\textit{Computing $s_2$: Random matrix product states}---For MPS, the $\ln(m)$ scaling is usually understood through bond-dimension additivity: namely, a superposition of $m$ MPS of bond dimension $\chi$ is an MPS of bond dimension $m\chi$, and thus $S \leq \ln(\chi) + \ln (m)$~\cite{Schollwock2011}.  We emphasize that this is a special case of our sub-maximally entangled class. A random MPS with $\chi \ll d^{|A|/2}$ has entanglement density $s_2 = |A|^{-1}\ln \bigl(d\chi/(d{+}1)\bigr)$, which for fixed $\chi$ vanishes as $|A|\to\infty$ (area law). Equivalently, the orthogonality rate $c_\perp = \ln (d) - s_2$ tends to the maximal value $\ln (d)$, so the component RDMs are almost perfectly orthogonal and the $\ln(m)$ scaling is immediate. 

The Schmidt spectrum of a random MPS sharpens this into a closed form. Defining rescaled coefficients $\mu_i = \sqrt{\chi}\lambda_i$, the spectrum collapses, for large $\chi$, onto a Marchenko--Pastur (MP) distribution~\cite{MarchenkoPastur1967}. The MP distribution describes the eigenvalues $x = \mu^2$
of Wishart-type random matrices. For aspect ratio $\gamma$
and variance parameter $\sigma^2$, it has density
\begin{equation}
\rho_{\mathrm{MP}}(x)
= \frac{\sqrt{(b-x)(x-a)}}{2\pi\,\sigma^2\gamma\,x},
\quad x \in [a,b],
\label{eq:mp_density}
\end{equation}
with $a = \sigma^2(1{-}\sqrt{\gamma})^2$ and $b = \sigma^2(1{+}\sqrt{\gamma})^2$~\cite{MarchenkoPastur1967}. The MPS simulation data shown in Fig.~\ref{fig:mp_multid} is fitted with variance $\sigma^2 = 1$ and aspect ratio $\gamma = 1/d$. 

The aspect ratio reflects the $\chi \times d\chi$ isometry shape of each site tensor, with the half-chain spectrum being governed by products of such random rectangular matrices~\cite{Collins2013}. The purity is the spectral fourth moment, with $\chi$ Schmidt values $\mathrm{Tr}\rho_A^2=\sum_i\lambda_i^4=\chi^{-1}\langle\mu^4\rangle$. As such, the 2nd Rényi entropy reads
\begin{equation}
S_2 = \ln\chi - \ln\langle\mu^4\rangle = \ln\!\frac{d\,\chi}{d+1},
\label{eq:s2_mps_exact}
\end{equation}
using the MP fourth moment $\langle\mu^4\rangle = (d{+}1)/d$; confirmed numerically to high precision for $d = 2, 3, 4$.  The inset of Fig.~\ref{fig:scaling} shows the resulting $\ln(m)$ scaling for MPS superpositions under a total bond-dimension cutoff, including the predicted plateau at $S_2 = \ln(d\chi_{\max}/(d{+}1))$ once $m\chi$ exceeds $\chi_{\max}$.

\textit{The maximal class: intensive relaxation onto Haar} -- When $c_\perp \to 0$, the purity tensor is no longer diagonally dominant, and the relaxation changes character. The B-side (``swap'') purity-tensor element switched off by one-sided orthogonality now survives, and tracking it yields exact, $N$-independent laws.  We quote the key results here and defer the derivations to Appendices~\ref{sec:maximal} and~\ref{sec:unfiltered}.

For {maximally entangled} components of the RDM (e.g., rank-deficit zero stabilizer states), superposition produces a genuine one-bit collapse from the maximal value at $m=1$ onto the Haar value of the 2nd Rényi entropy. Namely,
\begin{equation}
\Delta S_2(m) = -\ln\!\bigl(2 - \tfrac1m\bigr) \;\xrightarrow{m\to\infty}\; -\ln (2) ,
\label{eq:filtered}
\end{equation}
a one-bit ($\ln 2$) drop at a parameter-free rate $\sim 1/m$, independent of $N$.

For \emph{generic} stabilizer components, the relaxation is instead a pure fluctuation effect.  The purity is a two-copy quantity and random stabilizers form a $3$-design~\cite{Webb2016,Zhu2017}, so its ensemble mean is pinned to the Haar value at every $m$ 
(i.e., $\bigl\langle \mathrm{Tr}\rho_A^2 \bigr\rangle_{\mathrm{stab}}
= \bigl\langle \mathrm{Tr}\rho_A^2 \bigr\rangle_{\mathrm{Haar}}\quad \forall \,m$), 

and the entire $m$-dependence of $\langle S_2\rangle$ is a Jensen gap $J(m)$~\cite{Jensen1,Jensen2,Abramovich2016Dec} sourced by the integer rank-deficit fluctuations $\delta$, with $J(1)\to(1-\langle\delta\rangle)\ln (2)\approx0.10$ and $J(m\geq2)\approx\mathrm{Var}(2^\delta)/(8m^{3})$.  Since the pinned mean cancels in the difference, the gain takes the same $\Delta S$ form (for $m\geq2$) as the other classes,
\begin{equation}
\Delta S_2(m) = J(m)-J(1) \approx \frac{\mathrm{Var}(2^\delta)}{8\,m^{3}} - (1-\langle\delta\rangle)\ln (2) ,
\label{eq:jensen}
\end{equation}
which is again $N$-independent (rank-deficit statistics and closed-form moments in Appendix~\ref{sec:unfiltered}).  In contrast to the $+\ln (m)$ growth of the sub-maximal class, here $\Delta S<0$. Superposition removes the single-component gap $J(1)$ and moves the entanglement \emph{down} onto the Haar-typical $s_2$, essentially complete by $m=2$ with a $1/m^3$ tail.

\textit{Conclusions.}-- 
We have shown that the enhancement of entanglement via superpositions of random states depends on whether the components of the reduced density matrices have
sub-maximal 2nd Rényi entanglement entropy density $s_2$. For each random ensemble from the sub-maximally entangled class we consider, the input $s_2$ is a random-matrix computation: the Wachter (arcsine) law in the case of Gaussian covariance blocks, and the Marchenko-Pastur (MP) law with $\gamma = 1/d$ for MPS Schmidt spectra. 

The hierarchy of state families (i.e., MPS at low bond dimension, Gaussians at intermediate density, and Haar and stabilizer states at the "top") collapses onto the single axis $s_2/\ln (2) \in [0,1]$, with the two endpoint behaviors now both quantitative. There is an extensive $\ln (m)$ growth below the maximal value of $\ln(2)$ and intensive, $N$-independent relaxation laws Eqns.~\eqref{eq:filtered}--\eqref{eq:jensen} near the Haar limit. The maximally entangled class is ensemble-agnostic, with the one-bit collapse Eqn.~\eqref{eq:filtered} using only the unitarity of the connecting maps. The Clifford ensemble differs only by a Jensen gap.

\begin{acknowledgments}
{\it Acknowledgements} -- We thank Ben Lehany-Fee, David Murtagh, and Joost Slingerland for useful discussions. DQ is supported by the Central Bank of Ireland's Academic and Professional Training Scheme. As of May 1st 2026, JTH has been funded by a Pathway Grant from Taighde \'Eireann--Research Ireland, and is employed by Maynooth University. Specifically, this publication has emanated from research conducted with the financial support of Taighde \'Eireann--Research Ireland, Grant Number 24/PATH-S/12630. For the purpose of Open Access, the authors have applied a CC BY public copyright licence to any Author Accepted Manuscript (AAM) version arising from this submission. The AAM will be available for free on the arXiv (a free distribution service and an open-access archive) upon completion of peer review. The original manuscript will also be available for free on the arXiv prior to completion of peer review.
\end{acknowledgments}

\appendix

\begin{center}
    {\bf Appendices}
\end{center}

\setcounter{secnumdepth}{3}

\section{The purity tensor: incoherent mixtures, orthogonal support, and the entropy of superpositions}
\label{app:purity_tensor}

The entanglement of an $m$-component superposition is carried entirely by the \emph{purity tensor},
\begin{align}
\mathcal{T}_{ijk\ell}&=\mathrm{Tr}\!\big[\mathrm{Tr}_B(\ket{\psi_i}\!\bra{\psi_j})\,\mathrm{Tr}_B(\ket{\psi_k}\!\bra{\psi_\ell})\big],
\\
\mathrm{Tr}(\rho_A^2)&=\sum_{ijk\ell}\lambda_i\lambda_j^*\lambda_k\lambda_\ell^*\,\mathcal{T}_{ijk\ell}.
\label{eq:T_app}
\end{align}
This appendix sets out its structure: which index patterns survive the average, the distinct physical roles that orthogonality on the two subsystems plays in collapsing it onto the diagonal, and the entropy of the resulting orthogonal mixture, including its dependence on the weight distribution.

\subsection{Surviving index patterns}
\label{app:patterns}
The quadruple sum in Eq.~\eqref{eq:T_app} reduces to a diagonal under either of two averages. Over a random ensemble of independent components $\mathbb{E}\ket{\psi_a}=0$, so $\mathbb{E}[\mathcal{T}_{ijk\ell}]$ vanishes unless every component enters once as a ket and once as a bra; the only such pairings are the \emph{direct} ($i{=}j,\,k{=}\ell$) and the \emph{swap} ($i{=}\ell,\,k{=}j$). These are the two elements of the two-replica permutation commutant $\mathfrak{S}_2=\{1,\mathrm{SWAP}\}$ that organises random matrix-product-state and unitary-design replica averages~\cite{garnerone2010a,garnerone2010b,haferkamp2021}. 

For fixed states carrying independent random phases $\lambda_a=e^{i\varphi_a}|\lambda_a|$, these same two pairings are the only ones whose phases survive. They carry distinct objects,
\begin{align}
\mathcal{T}_{iikk}&=\mathrm{Tr}\!\big(\rho_A^{(i)}\rho_A^{(k)}\big),
\\
\mathcal{T}_{ikki}&=\big\|\mathrm{Tr}_B\ket{\psi_i}\!\bra{\psi_k}\big\|_{\mathrm{HS}}^2=\mathrm{Tr}\!\big(\rho_B^{(i)}\rho_B^{(k)}\big),
\label{eq:direct_swap}
\end{align}
the direct term the overlap of two $A$-reduced states and the swap term the overlap of two $B$-reduced states, the last equality being the identity $\|\mathrm{Tr}_B\ket{\psi}\!\bra{\phi}\|_{\mathrm{HS}}^2=\mathrm{Tr}(\rho_\psi^B\rho_\phi^B)$ of Appendix~\ref{sec:proof}. The fully diagonal entry $i{=}j{=}k{=}\ell$ is the single-component purity $\mathcal{T}_{iiii}=\mathrm{Tr}\big((\rho_A^{(i)})^2\big)=e^{-S_2^{(i)}}$, shared by both pairings.

\subsection{The one-sided-orthogonality criterion}
\label{app:criterion}
For an ensemble whose average RDM is maximally mixed, $\bar\rho^A=I/2^{|A|}$ (Schur's lemma, whenever a group acts irreducibly on the symmetry-restricted sector), both off-diagonal terms of Eq.~\eqref{eq:direct_swap} collapse to the universal cross-overlap $2^{-|A|}$: by independence $\mathbb{E}[\mathrm{Tr}(\rho_\psi^A\rho_\phi^A)]=\mathrm{Tr}((\bar\rho^A)^2)=2^{-|A|}$, fixed by the Hilbert-space dimension alone. Set against the diagonal self-purity $e^{-s_2|A|}$, this gives the cross-to-self ratio $C_0\,e^{-c_\perp|A|+o(|A|)}$ of Eq.~\eqref{eq:orthogonality}, which vanishes iff $s_2<\ln 2$. The ratio is the natural orthogonality measure: it weighs the cross-overlap of two distinct RDMs against one RDM's own purity, $\mathrm{Tr}(\rho_\psi^A\rho_\phi^A)/\mathrm{Tr}((\rho^A)^2)$. The cross-overlap $2^{-|A|}$ is itself never zero, but a vanishing ratio makes it negligible beside the self-purity $e^{-s_2|A|}$, so the two component RDMs are asymptotically orthogonal, $\mathrm{Tr}(\rho_\psi^A\rho_\phi^A)\ll\mathrm{Tr}((\rho^A)^2)$. This is the general one-sided-orthogonality criterion: it turns on the single number $s_2$, with the $\mathcal{O}(1)$ prefactor $C_0$ absorbing any conserved charge that leaves $\bar\rho^A$ maximally mixed only within sectors. The criterion applies to every sub-maximal class; we verify it rigorously for the fermionic Gaussian class, where both $\bar\rho^A$ and $c_\perp$ are computable, in Appendix~\ref{sec:proof}.

\subsection{The two sides of orthogonality}
\label{app:two_faces}
The reduced state of the superposition splits into an incoherent part and a coherent one,
\begin{equation}
\rho_A=\sum_i p_i\,\rho_A^{(i)}+\sum_{i\neq k}\lambda_i\lambda_k^*\,\mathrm{Tr}_B\ket{\psi_i}\!\bra{\psi_k},
\label{eq:rhoA_decomp}
\end{equation}
with $p_i=|\lambda_i|^2$; the first sum is the incoherent mixture, the second the coherences. The direct and swap pairings of $\mathcal{T}$ (Eq.~\eqref{eq:direct_swap}) control these two parts, and orthogonality on the two subsystems does \emph{different} work.

\paragraph{$B$-side orthogonality: decoherence into a mixture.}
Each coherence in Eq.~\eqref{eq:rhoA_decomp} is a partial trace $\mathrm{Tr}_B\ket{\psi_i}\!\bra{\psi_k}$ whose magnitude is the swap overlap $\|\mathrm{Tr}_B\ket{\psi_i}\!\bra{\psi_k}\|_{\mathrm{HS}}^2=\mathrm{Tr}(\rho_B^{(i)}\rho_B^{(k)})$ of Eq.~\eqref{eq:direct_swap}. One-sided orthogonality on $B$ sends this to zero, so the coherences vanish and $\rho_A$ collapses to the \emph{incoherent mixture} $\sum_i p_i\rho_A^{(i)}$: tracing out $B$ records which component is present and decoheres the superposition.

\paragraph{$A$-side orthogonality: orthogonal support.}
Whether that mixture is strongly entangled then turns on how the $\rho_A^{(i)}$ sit relative to one another. One-sided orthogonality on $A$, $\mathrm{Tr}(\rho_A^{(i)}\rho_A^{(k)})\to0$ (the direct overlap), makes the component states \emph{orthogonally supported}: the $m$ components occupy disjoint subspaces of the $A$ Hilbert space.

Only when \emph{both} hold---two-sided orthogonality---is $\rho_A$ a mixture of orthogonally supported states, and only then does the entanglement gain a full $\ln m$ (Sec.~\ref{sec:entropy}). $B$-side orthogonality alone gives an incoherent mixture of \emph{overlapping} states, with a reduced mixing gain; $A$-side alone leaves residual coherences and no universal closed form. At a balanced cut the two conditions coincide, since $S(\rho_A)=S(\rho_B)$ ties $c_\perp>0$ on the two subsystems together.

\subsection{Entropy of an orthogonal mixture}
\label{sec:entropy}

Consider an incoherent mixture
$\rho = \sum_i p_i\rho_i$ whose components satisfy
$\mathrm{Tr}(\rho_i\rho_j) = 0$ for $i\neq j$.
This is the two-sided-orthogonal mixture of Sec.~\ref{app:two_faces}.

\textit{Von Neumann entropy.}
Using the concavity of $S$ and the orthogonality condition,
$S(\rho) = H(\{p_i\}) + \sum_i p_i S(\rho_i)$,
where $H$ is the Shannon entropy of the weights.
For equal weights: $S = \ln(m) + \langle S(\rho_i)\rangle$.

\textit{$2$nd R\'{e}nyi entanglement entropy.}
The purity of the mixture is
$\mathrm{Tr}(\rho^2)
= \sum_{i,j} p_i p_j\,\mathrm{Tr}(\rho_i\rho_j)
= \sum_i p_i^2\,\mathrm{Tr}(\rho_i^2)$,
where the second equality uses
$\mathrm{Tr}(\rho_i\rho_j) = 0$ for $i \neq j$.
Hence $S_2(\rho) = -\ln\sum_i p_i^2\,e^{-S_2(\rho_i)}$.
This does not decompose as mixing entropy plus average.
For equal weights:
$S_2 = \ln(m) - \ln\langle e^{-S_2(\rho_i)}\rangle$.
When component entropies concentrate (as occurs for large systems),
$\langle e^{-S_2}\rangle \approx e^{-\langle S_2\rangle}$,
recovering $S_2(m) \approx \langle S_2(1)\rangle + \ln(m)$.

The number of components with mutually orthogonal supports is
bounded by $m \lesssim d/r = e^{S_{\max} - S}$:
when $S \ll S_{\max}$, exponentially many orthogonal states fit;
when $S \approx S_{\max}$, at most $O(1)$ can.

\subsection{Non-equal weights}
\label{sec:nonequal}

The results in the main text focus on equal-weight superpositions
$|\lambda_k|^2 = 1/m$.
A natural question is whether the $\ln(m)$ scaling persists for general coefficient
distributions.

For arbitrary weights $\{p_k = |\lambda_k|^2\}$ summing to unity,
the purity of the reduced state under RDM orthogonality is
\begin{equation}
\mathrm{Tr}(\rho_A^2)
\approx \sum_k p_k^2\,e^{-S_2^{(k)}},
\label{eq:purity_general}
\end{equation}
and the $2$nd R\'{e}nyi entropy becomes
$S_2 \approx -\ln\!\sum_k p_k^2 - \ln\langle e^{-S_2^{(k)}}\rangle$,
where the first term is the collision entropy $H_2(\{p_k\})$
rather than $\ln(m)$.
The von~Neumann entropy similarly gives
$S \approx H(\{p_k\}) + \sum_k p_k S^{(k)}$,
where $H$ is the Shannon entropy.

Equal weights maximise both $H$ and $H_2$ at $\ln(m)$: any other
weight distribution produces less mixing entropy.
The physical mechanism is a \textit{fourth-moment effect}:
the purity $\mathrm{Tr}(\rho_A^2)$ depends on
$\sum_k p_k^2 = \|\mathbf{p}\|_2^2$, which is minimised
(and hence $S_2$ is maximised) when all weights are equal.

For random complex coefficients $c_k \sim \mathcal{CN}(0,1)$
normalised to $\sum|c_k|^2 = 1$, the weights $p_k = |c_k|^2/\sum|c_j|^2$
follow a symmetric Dirichlet$(1,\ldots,1)$ distribution.
The expected Shannon entropy is exactly
\begin{equation}
\mathbb{E}[H(\{p_k\})]
= \psi(m{+}1) - \psi(2)
= H_m - 1,
\label{eq:shannon_dirichlet}
\end{equation}
where $\psi$ is the digamma function and
$H_m = \sum_{k=1}^m 1/k$ is the $m$-th harmonic number.
The deficit relative to the equal-weight value $\ln(m)$ is
$\Delta_{\mathrm{vN}}(m) = \ln(m) - (H_m - 1)$,
which vanishes at $m = 1$, grows monotonically, and saturates
at $1 - \gamma_E \approx 0.423$ as $m \to \infty$
($\gamma_E \approx 0.577$ is the Euler--Mascheroni constant).
The two curves therefore agree at $m = 1$, fan apart for
small~$m$, and become parallel (both growing as $\ln m$)
only asymptotically.

For the $2$nd R\'{e}nyi entropy, no analogous closed form exists.
The expected collision entropy
$\mathbb{E}[H_2] = \mathbb{E}[-\ln\sum p_k^2]$
is well approximated by the asymptotic expansion
\begin{equation}
\mathbb{E}[H_2]
= \ln\!\frac{m{+}1}{2}
+ \frac{m{-}1}{2(m{+}2)(m{+}3)}
+ O(1/m^2),
\label{eq:collision_dirichlet}
\end{equation}
where the leading term is the Jensen bound
$-\ln\mathbb{E}[\sum p_k^2] = \ln\!\tfrac{m{+}1}{2}$
(using the exact Dirichlet moment
$\mathbb{E}[\sum p_k^2] = 2/(m{+}1)$)
and the correction arises from the variance of $\sum p_k^2$.
The deficit $\Delta_{R2}(m) = \ln(m) - \mathbb{E}[H_2]$
vanishes at $m = 1$ and saturates at $\ln 2 \approx 0.693$
for large $m$---larger than the von~Neumann deficit
$1 - \gamma_E \approx 0.423$ because the collision entropy
is more sensitive to weight inhomogeneity.

For the $2$nd R\'{e}nyi entropy, random phases additionally cause
partial cancellation of interference terms.
When the component states are nearly orthogonal
($\braket{G_i|G_j} \approx 0$ for $i\neq j$), random phases in
the coefficients suppress the off-diagonal purity tensor elements
$\mathcal{T}_{ijk\ell}$ with $i \neq j$ or $k \neq \ell$,
driving the RDM towards an incoherent mixture even without
invoking RDM orthogonality.
This means that for random-weight superpositions, the $\ln(m)$
scaling receives contributions from two complementary effects:
the phase cancellation of interference terms (which operates
even without the rank gap), and the RDM orthogonality (which
operates even for equal weights).

\section{One-sided orthogonality of random Gaussian states}
\label{sec:proof}

We verify the general one-sided-orthogonality criterion of Appendix~\ref{app:criterion} for the fermionic Gaussian class, where both ingredients it calls for can be computed in closed form. The first is the average reduced density matrix, which fixes the cross-overlap in the numerator of the orthogonality ratio; the second is the typical purity, which fixes the self-overlap in the denominator and with it the rate $c_\perp$. Together they show that the cross-to-self ratio decays exponentially in $|A|$.

\subsection{The maximally mixed average RDM}
The average RDM $\bar\rho^A=\mathbb{E}[\rho_\psi^A]$ is fixed entirely by symmetry. Haar-random Gaussian states form the orbit $\{\hat U_O\ket{0}:O\in O(2N)\}$ of the vacuum under Bogoliubov transformations, and the representation of $\mathrm{Spin}(2N)$ on the even-parity Fock space $\mathcal H_+$ (dimension $2^{N-1}$) is irreducible~\cite{Georgi}. Schur's lemma therefore pins the ensemble average to the identity on that sector, $\mathbb{E}[\ket\psi\!\bra\psi]=P_+/2^{N-1}$. Tracing over $B$ and using $P_+=P_A^+\otimes P_B^+ + P_A^-\otimes P_B^-$,
\begin{equation}
\bar\rho^A=\mathrm{Tr}_B\!\left(\frac{P_+}{2^{N-1}}\right)
=\frac{(P_A^++P_A^-)\,2^{N-|A|-1}}{2^{N-1}}
=\frac{I_{2^{|A|}}}{2^{|A|}},
\end{equation}
since $P_A^++P_A^-=I_{2^{|A|}}$. The average RDM is thus maximally mixed, and by independence of $\ket\psi$ and $\ket\phi$ the cross-overlap sits at the universal floor $\mathbb{E}[\mathrm{Tr}(\rho_\psi^A\rho_\phi^A)]=\mathrm{Tr}((\bar\rho^A)^2)=2^{-|A|}$. The same symmetry argument covers the number-conserving case, changing only the prefactor: for Slater determinants with $M$ particles, $U(N)$ invariance makes $\bar\rho^A$ maximally mixed within each particle-number sector,
\begin{equation}
\begin{aligned}
\bar\rho^A&=\bigoplus_{n_A}\frac{\binom{N-|A|}{M-n_A}}{\binom{N}{M}}\,\frac{I_{\binom{|A|}{n_A}}}{\binom{|A|}{n_A}},\\[2pt]
\mathrm{Tr}\big((\bar\rho^A)^2\big)&=\sum_{n_A}\frac{\binom{|A|}{n_A}\binom{N-|A|}{M-n_A}^2}{\binom{N}{M}^2}\sim\frac{2}{\sqrt{3}}\,2^{-|A|}
\end{aligned}
\end{equation}
at half filling ($M=N/2$). This extra $\mathcal O(1)$ factor leaves the exponential decay untouched.

\subsection{The typical purity}
The self-overlap follows from Peschel's factorization~\cite{peschel}. In the eigenbasis of the covariance subblock $\Gamma^A$, the Gaussian RDM is a product over modes,
\begin{equation}
\rho^A=\bigotimes_{k=1}^{|A|}
\left[\frac{1+\epsilon_k}{2}\ket{0_k}\!\bra{0_k}
+\frac{1-\epsilon_k}{2}\ket{1_k}\!\bra{1_k}\right],
\end{equation}
where $\pm i\epsilon_k$ are the eigenvalues of $\Gamma^A$ and $\epsilon_k\in[0,1]$. Each mode contributes independently, so
\begin{align}
\mathrm{Tr}((\rho^A)^2)
&=\prod_{k=1}^{|A|}\!\left[
\left(\frac{1{+}\epsilon_k}{2}\right)^{\!2}
\!+\left(\frac{1{-}\epsilon_k}{2}\right)^{\!2}\right]\notag\\
&=2^{-|A|}\prod_{k=1}^{|A|}(1+\epsilon_k^2),
\label{eq:purity_product}
\end{align}
or equivalently $2^{-|A|}[\det(I_{2|A|}+(\Gamma^A)^\top\Gamma^A)]^{1/2}$, since the positive matrix $(\Gamma^A)^\top\Gamma^A$ has eigenvalues $\epsilon_k^2$ (each doubly degenerate). This product exceeds the floor $2^{-|A|}$ by an extensive factor whenever the $\epsilon_k$ are bounded away from $1$, and that is exactly what a random Gaussian state provides: the $\{\epsilon_k^2\}$ form a Jacobi (MANOVA) ensemble~\cite{bhk,wachter}, and in the limit $N\to\infty$ at fixed $\alpha=|A|/N$ their empirical distribution converges to the Wachter law $\rho_W(\epsilon)$ supported on $[\epsilon_-,\epsilon_+]\subset(0,1)$. The per-mode rate $c_\perp=|A|^{-1}\sum_k\ln(1+\epsilon_k^2)$ is a linear statistic that self-averages~\cite{johansson} (with variance $O((\ln|A|)^2/|A|^2)$) to
\begin{equation}
c_\perp=\int_0^1\ln(1+\epsilon^2)\,\rho_W(\epsilon)\,d\epsilon>0 ,
\end{equation}
so the typical purity is $\mathbb{E}[\mathrm{Tr}((\rho_\psi^A)^2)]=2^{-|A|}e^{c_\perp|A|+o(|A|)}$.

\subsection{Exponential decay of the ratio}
Dividing the cross-overlap by the self-purity gives the one-sided-orthogonality ratio for the Gaussian class,
\begin{equation}
\frac{\mathbb{E}[\mathrm{Tr}(\rho_\psi^A\rho_\phi^A)]}
     {\mathbb{E}[\mathrm{Tr}((\rho_\psi^A)^2)]}
=C_0\,e^{-c_\perp|A|+o(|A|)}\to 0 ,
\label{eq:main_theorem}
\end{equation}
with $C_0=1$ for BdG states and $C_0\to 2/\sqrt{3}$ as $|A|\to\infty$ for number-conserving states at half filling. The $B$-side orthogonality follows by the identical argument, since $\|\mathrm{Tr}_B(\ket\psi\!\bra\phi)\|_{\mathrm{HS}}^2=\mathrm{Tr}(\rho_\psi^B\rho_\phi^B)$ obeys the same two estimates. This is the rigorous Gaussian instance of the general criterion; i.e., the numerator is fixed by Hilbert-space dimension alone, the denominator sits an extensive factor $e^{c_\perp|A|}$ above it, and their vanishing ratio leaves exponentially many orthogonal directions for independent components to occupy.

\subsection{Explicit $c_\perp$ at equal bipartition}
At $\alpha = 1/2$, for both BdG and number-conserving states
at half filling, the Wachter distribution reduces to the arcsine
form $\rho_\lambda(\lambda) = 2/[\pi\sqrt{1-\lambda^2}]$
on $[0,1]$~\cite{bhk,bianchi}.
Substituting $\lambda = \sin\theta$:
\begin{equation}
c_\perp
= \frac{2}{\pi}\int_0^{\pi/2}
  \ln(1+\sin^2\theta)\,d\theta.
\end{equation}
Writing $1+\sin^2\theta = \cos^2\theta + 2\sin^2\theta$
and applying the standard identity
$\int_0^{\pi/2}\ln(a^2\cos^2\theta + b^2\sin^2\theta)\,d\theta
= \pi\ln\bigl(\tfrac{a+b}{2}\bigr)$
with $a=1$, $b=\sqrt{2}$:
\begin{equation}
c_\perp
= 2\ln\!\left(\frac{1+\sqrt{2}}{2}\right)
= \ln\!\left(\frac{3+2\sqrt{2}}{4}\right)
\approx 0.376.
\end{equation}

\subsection{Comparison with full Haar-random states}
For states drawn from the full Fock space, the average RDM
is also $I/2^{|A|}$, so the numerator is the same.
The denominator, however, is barely above the floor:
using the 2-replica method, the typical purity is
$(d_A + d_B)/(d_Ad_B + 1) \approx 2^{-|A|} + 2^{-(N-|A|)}$.
At equal bipartition this gives
$\langle S_2\rangle = |A|\ln 2 - \ln 2$ and hence
$c_\perp^{\mathrm{Haar}} = \ln 2/|A|$: nonzero at any finite
$|A|$, but vanishing as $|A| \to \infty$, so $s_2 \to \ln 2$
saturates the ceiling in the thermodynamic limit.
The product $c_\perp |A| = \ln 2$ is independent of system
size, so the cross-to-self ratio sits at $1/2$ regardless of
$|A|$ (no orthogonality).
This is the sense in which Haar states belong to the
$s_2 = \ln d$ class: the saturation is asymptotic,
and the finite-$|A|$ deficit ($\ln 2$ for $S_2$; $\tfrac12$ for the
von~Neumann Page value) is an intensive
correction rather than the extensive gap $c_\perp = \Theta(1)$
needed for RDM orthogonality.
Gaussian states, by contrast, have
$S_2 \approx 0.317\,|A|$ ($46\%$ of maximum), giving
$c_\perp \approx 0.376$ independent of $|A|$ and leaving
exponentially many orthogonal directions for independent
states to occupy.

%======================================================================
\section{Pfaffian Evaluation of the Purity Tensor}
\label{sec:boutin_bauer}
%======================================================================

The purity tensor elements $\mathcal{T}_{ijk\ell}$ for Gaussian
states are computed numerically using the Pfaffian method of
Boutin and Bauer~\cite{BoutinPRR2021}.
Each element is expressed as
$\mathcal{T}_{ijk\ell} = I_{ijk\ell}/(G_{ij}G_{k\ell})$,
where $G_{ij} = \braket{G_i|G_j}$ and $I_{ijk\ell}$ is a product
of four Pfaffians of matrices constructed from covariance-matrix
blocks.

For each pair $(\alpha,\beta)$, define:
\begin{align}
M_A^{\alpha\beta} &= \begin{pmatrix}
\Gamma_\alpha^{AA} & iI_{2|A|} \\
-iI_{2|A|} & \Gamma_\beta^{AA}
\end{pmatrix}, \\
M_B^{\alpha\beta} &= \begin{pmatrix}
\Gamma_\alpha^{BB} & iI_{2(N-|A|)} \\
-iI_{2(N-|A|)} & \Gamma_\beta^{BB}
\end{pmatrix},
\end{align}
and the coupling correction
$\Sigma_A^{\alpha\beta}
= C^{\alpha\beta}(M_B^{\alpha\beta})^{-1}(C^{\alpha\beta})^\top$,
where $C^{\alpha\beta}
= \mathrm{diag}(\Gamma_\alpha^{AB},\Gamma_\beta^{AB})$.
The auxiliary correction is
$\sigma_A^{\delta\gamma}
= -X(M_A^{\gamma\delta} + \Sigma_A^{\gamma\delta})^{-1}X$
with $X = \bigl(\begin{smallmatrix}-I & I \\ I & -I\end{smallmatrix}\bigr)$.

The Pfaffian product is then:
\begin{equation}
I_{ijk\ell}
= \frac{2^{N-|A|}}{8^N}\,
\mathrm{Pf}(M_B^{ij})\cdot\mathrm{Pf}(M_B^{k\ell})
\cdot\mathrm{Pf}(P^{jk})\cdot\mathrm{Pf}(Q^{ijk\ell}),
\end{equation}
where $P^{jk} = M_A^{jk} + \Sigma_A^{jk}$ and
$Q^{ijk\ell} = M_A^{ij} + \Sigma_A^{ij} + \sigma_A^{\ell k}$.
The total cost is $O(m^4 N^3)$.
For number-conserving (Slater) components the same elements reduce
to standard determinant formulas and the Pfaffian construction is
not required.

%======================================================================
\section{Maximally entangled components: the one-bit collapse}
\label{sec:maximal}
\label{app:filtered}
%======================================================================

This appendix and the next derive the two relaxation laws quoted in the
main text, Eqs.~\eqref{eq:filtered} and~\eqref{eq:jensen}.  Throughout we
take a balanced cut, $d_A = d_B = d = 2^{|A|}$, and equal weights
$p_j = |\lambda_j|^2$ with $\sum_j p_j = 1$.  The maximally entangled
(one-bit collapse) case is treated here, while the generic stabilizer (Jensen
gap) case is dealt with in Appendix~\ref{sec:unfiltered}.  We stress at the outset that
the calculation of this section uses only the \emph{unitarity} of the
connecting maps---it never invokes the stabilizer structure---so it holds
for any maximally entangled component, whether or not its connecting unitary is Clifford, and is in
that sense more general than the stabilizer-specific analysis that follows.

\subsection{Connecting-unitary reduction}
Any maximally entangled state has flat, full-rank Schmidt form and
can be written as $\ket{\psi} = (U_A \otimes U_B)\ket{\Omega}$, where
$\ket{\Omega} = d^{-1/2}\sum_k \ket{k}_A\ket{k}_B$ is a fixed
reference.  Using the ricochet identity
$(M \otimes I_B)\ket{\Omega} = (I_A \otimes M^{\mathsf T})\ket{\Omega}$,
the $A$-side unitary can be moved to the $B$ side,
$\ket{\psi} = (I_A \otimes V)\ket{\Omega}$ with
$V = U_B U_A^{\mathsf T}$.  Because every component shares the one
reference $\ket{\Omega}$, the superposition folds onto a single
operator:
\begin{equation}
\ket{\Psi_m} = (Y \otimes I_B)\ket{\Omega},
\qquad
Y = \sum_j \lambda_j V_j^{\mathsf T},
\end{equation}
and tracing out $B$ gives
\begin{equation}
\rho_A = \frac{Y Y^\dagger}{\mathcal{N}\,d},
\qquad
\mathcal{N} = \braket{\Psi_m|\Psi_m}
= \sum_{jk} \lambda_j^*\lambda_k\,
\tfrac1d \mathrm{Tr}(V_j^\dagger V_k).
\label{eq:YYdag}
\end{equation}
The entanglement spectrum is the normalised squared singular-value
spectrum of the weighted sum of unitaries $Y$.  For stabilizer
components the $V_j$ are Clifford unitaries; nothing below uses more
than their unitarity and (for the error estimate) the 2-design
property, so the result applies equally to Haar connecting
unitaries.

\subsection{Purity}
Expanding $YY^\dagger = \sum_{jk}\lambda_j\lambda_k^*(V_k^\dagger V_j)^{\mathsf T}$, the diagonal $j = k$ terms sum to $\sum_j|\lambda_j|^2\,I = I$ for normalised weights, leaving
\begin{equation}
\rho_A = \frac{I + X}{\mathcal{N}\,d},
\qquad
X = \sum_{j\neq k}\lambda_j\lambda_k^*(V_k^\dagger V_j)^{\mathsf T},
\end{equation}
the interference operator $X$ being built from pairs of distinct
components.  Squaring and tracing,
\begin{equation}
\mathrm{Tr}(\rho_A^2)
= \frac{\mathrm{Tr}\!\left[(I+X)^2\right]}{\mathcal{N}^2 d^2}
= \frac{d + 2\,\mathrm{Tr}\,X + \mathrm{Tr}(X^2)}{\mathcal{N}^2 d^2}.
\end{equation}
Now $\mathrm{Tr}\,X = d(\mathcal{N}-1)$ (from $\mathrm{Tr}(YY^\dagger) =
d\mathcal{N}$), and the norm departs from unity only through the
off-diagonal overlaps,
$\mathcal{N} = 1 + \sum_{j\neq k}\lambda_j^*\lambda_k\,
\tfrac1d\mathrm{Tr}(V_j^\dagger V_k)$, which for distinct (Clifford or
Haar) unitaries are $O(1/\sqrt d)$ with zero mean.  Thus
$\mathcal{N} = 1 + O(1/\sqrt d)$, the middle term is
$2\,\mathrm{Tr}\,X/d = 2(\mathcal{N}-1) = O(1/\sqrt d)$, and the prefactor
$\mathcal{N}^2 \to 1$, leaving
\begin{equation}
d\,\mathrm{Tr}(\rho_A^2) = 1 + \tfrac1d\,\mathrm{Tr}(X^2) + O(1/\sqrt d).
\end{equation}
The remaining trace,
$\mathrm{Tr}(X^2) = \sum_{j\neq k}\sum_{p\neq q}
\lambda_j\lambda_k^*\lambda_p\lambda_q^*\,
\mathrm{Tr}(V_q^\dagger V_p V_k^\dagger V_j)$.
The pairing $(p,q) = (k,j)$ is \emph{deterministic}:
$\mathrm{Tr}(V_j^\dagger V_k V_k^\dagger V_j) = \mathrm{Tr}(I) = d$
for any unitaries, contributing exactly
\begin{equation}
\tfrac1d\,\mathrm{Tr}(X^2)\big|_{\text{pairing}}
= \sum_{j\neq k} p_j p_k = 1 - \|p\|_2^2 .
\end{equation}
Every other index combination is a trace of a product of
\emph{distinct} random unitaries, which have zero mean and contribute
only $O(1/\sqrt d)$ in total, so the purity self-averages---it equals
its typical value, not merely its mean.

Putting the pieces together: only the deterministic pairing survives, so
$\tfrac1d\mathrm{Tr}(X^2) = (1 - \|p\|_2^2) + O(1/\sqrt d)$, and inserting
this into $d\,\mathrm{Tr}(\rho_A^2) = 1 + \tfrac1d\mathrm{Tr}(X^2)$ gives,
for any weight profile,
\begin{equation}
\mathrm{Tr}(\rho_A^2) = 2^{-|A|}\bigl(2 - \|p\|_2^2\bigr) + O(d^{-1/2}).
\end{equation}
For equal weights $\|p\|_2^2 = 1/m$ this is
\begin{equation}
\mathrm{Tr}(\rho_A^2) = 2^{-|A|}\Bigl(2 - \tfrac1m\Bigr)
+ O(d^{-1/2}),
\end{equation}
Taking the negative of the natural log of the above yields $\langle S_2(m)\rangle=(|A|-\log_2(2-1/m))\ln 2$, so the gain over the
single (maximal) component is $\Delta S(m)=-\ln(2-1/m)$, which is Eq.~\eqref{eq:filtered}.
Because the quantity inside the logarithm does not fluctuate at leading order, no Jensen
correction arises and the annealed and quenched entropies coincide: the law is exact at
$m=1$ (where $\Delta S=0$) and relaxes onto the Haar value $(|A|-1)\ln 2$ as $m\to\infty$,
a one-bit drop $\Delta S\to-\ln (2)$ approached at rate $\sim 1/(2m)$ nats.

\section{Generic stabilizer superpositions: the Jensen gap}
\label{sec:unfiltered}

\subsection{The rank deficit}
A stabilizer state is flat on its Schmidt support, with purity
$\mathrm{Tr}(\rho_A^2) = 2^{-(|A|-\delta)}$, where the non-negative integer
$\delta = |A| - S_2/\ln (2)$ is the \emph{rank deficit}---the number of bits
by which the entanglement falls short of the maximal value
$S_2 = |A|\ln 2$.  Equivalently, $\delta$ counts the independent stabilizer generators that
can be supported entirely on $A$, each one costing a bit of entanglement
across the cut; the state is maximally entangled ($\delta = 0$) precisely
when no such $A$-local generator exists~\cite{Fattal2004}.  For random stabilizer states the
distribution of $\delta$ is $N$-independent up to $O(2^{-N})$ corrections,
with $\langle\delta\rangle \approx 0.86$ and a fraction
$\prod_{i\geq1}(1-2^{-i}) \approx 0.29$ maximally
entangled~\cite{SmithPRA2006,DahlstenQIC2006}.

\subsection{Group-membership probabilities}
Two facts about a uniformly random stabilizer group $\mathcal{G}$ drive the
moment count below.  Recall that a stabilizer group is a set of $2^N$
mutually commuting Paulis closed under multiplication---the identity
together with $2^N-1$ nontrivial elements.  No nontrivial Pauli is
privileged over any other: any one can be relabelled into any other by a
change of basis that preserves all commutation relations (a Clifford
operation), and the ensemble of stabilizer groups is unchanged by such
relabellings.  Every nontrivial Pauli therefore lies in the same fraction
of groups, and that fraction is simply the number a group holds divided by
the number that exist.  Of the $2^{2N}-1$ nontrivial Paulis on $N$ qubits,
each group contains $2^N-1$, so
\begin{equation}
\Pr(P\in\mathcal{G}) = \frac{2^N-1}{2^{2N}-1} = \frac{1}{2^N+1}.
\label{eq:prob_single}
\end{equation}
For two Paulis, condition on $P\in\mathcal{G}$ and ask whether $Q$ joins
too.  Since $\mathcal{G}$ is abelian, $Q$ can only belong if it commutes
with $P$; an anticommuting $Q$ is excluded outright.  Using that any two
Paulis either commute or anticommute, and that exactly half of all Paulis
commute with a fixed nontrivial one, of those $2^{2N-1}$ commuting Paulis
discard the identity and $P$ itself, leaving $2^{2N-1}-2$ candidates for an
independent commuting $Q$; of these, $\mathcal{G}$ holds $2^N-2$ (its $2^N$
elements minus the identity and $P$).  The same ``nothing is privileged''
counting then gives
\begin{equation}
\Pr(Q\in\mathcal{G}\mid P\in\mathcal{G})
= \frac{2^N-2}{2^{2N-1}-2}
= \frac{1}{2^{N-1}+1},
\end{equation}
and multiplying by Eq.~\eqref{eq:prob_single},
\begin{equation}
\Pr(P,Q\in\mathcal{G}) = \frac{1}{(2^N+1)(2^{N-1}+1)}
\label{eq:prob_pair}
\end{equation}
for a commuting independent pair (and zero for an anticommuting one).

\subsection{Design pinning}
The purity of the superposition contains each component projector to
at most second order, so its ensemble mean is a 2-design quantity.
Random stabilizer states form an exact 3-design~\cite{Webb2016,Zhu2017},
which gives the design pinning: the mean purity equals the Haar
value exactly, for every $m$.  (Consistently, the single-state mean purity
$\langle \mathrm{Tr}\rho_A^2\rangle = 2^{-|A|}\langle 2^\delta\rangle$
reduces to the Haar value $2\cdot 2^{-|A|}$ once the deficit moments
computed below are inserted.)
Only the \emph{variance} of the purity probes the 4-design property that
stabilizer states fail.

\begin{figure*}[t]
\includegraphics[width=15cm]{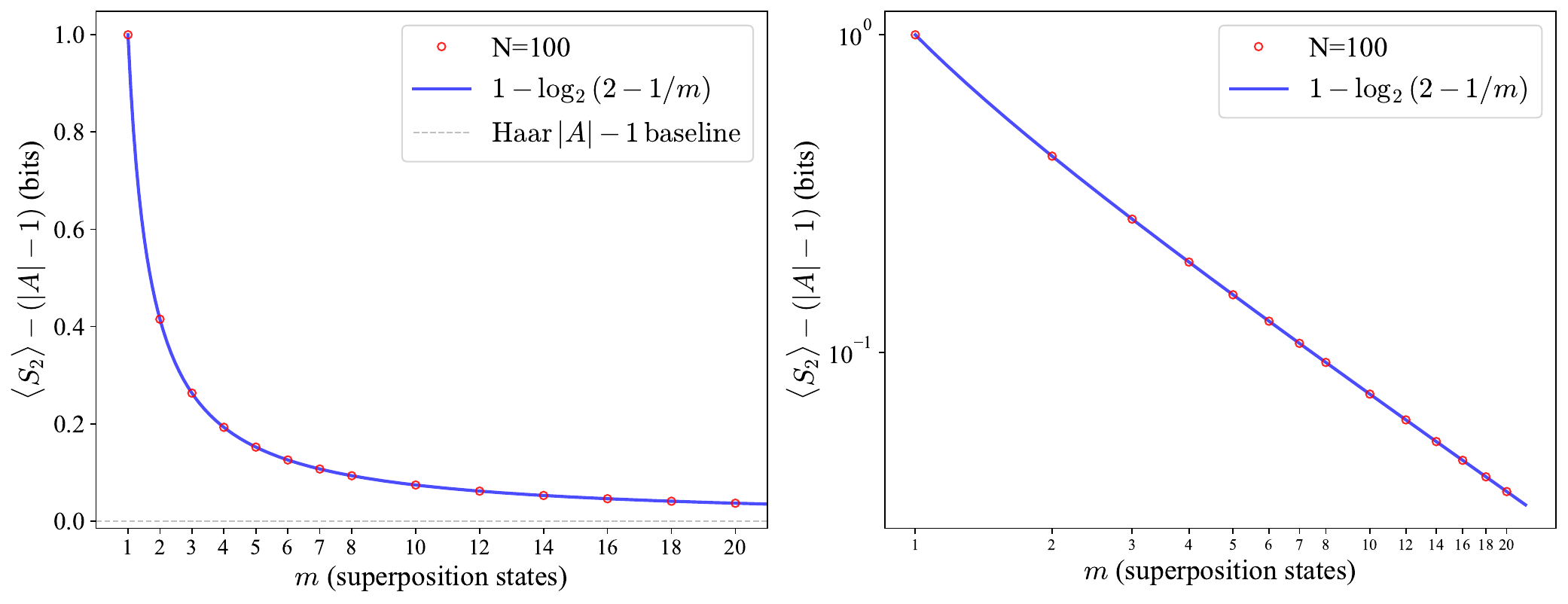}\\
\includegraphics[width=15cm]{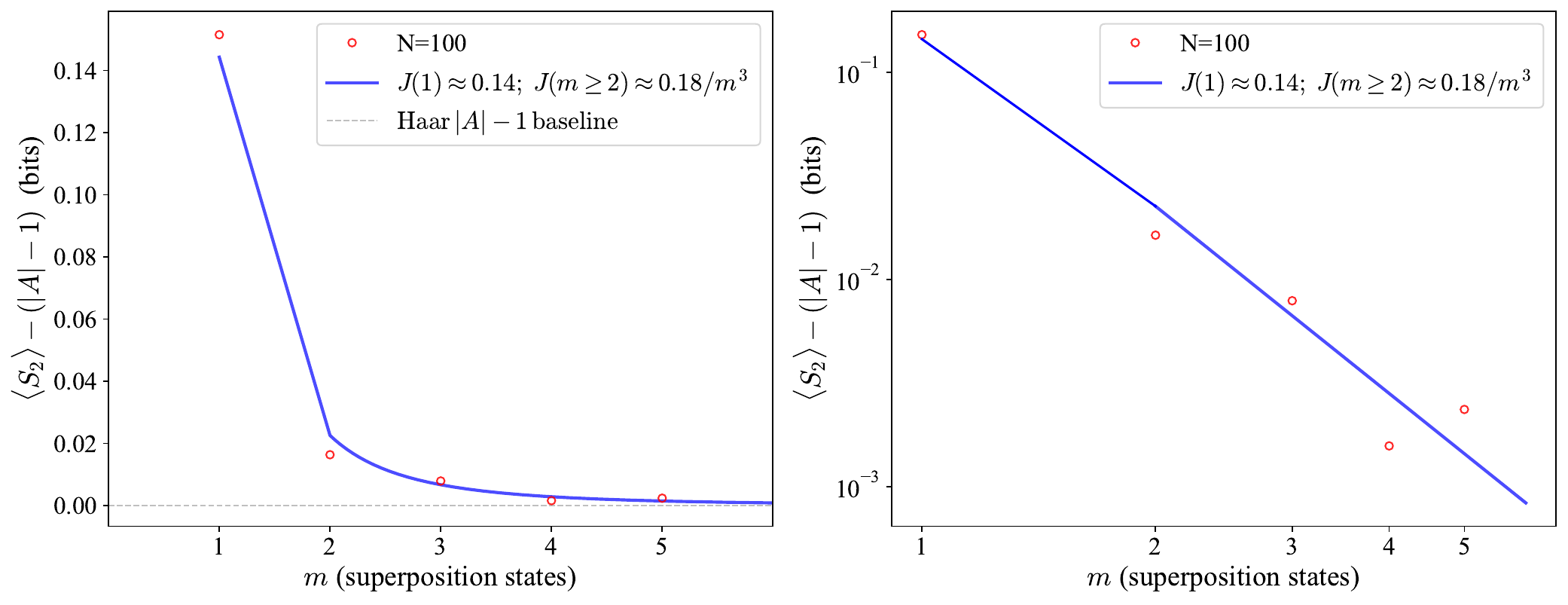}
\caption{(Top) Numerical calculations of the maximally entangled $\sim 1/m$ relaxation of $\langle S_2 \rangle$ to the Haar average. (Bottom) The relaxation of the fully random Clifford ensemble $~1/m^3$ relaxation to the Haar average. The log-log scalings of both $\langle S_2 \rangle$ deviations are shown on the right-hand side.  Numerical calculations completed at large system sizes using the stabilizer representation of \cite{BravyiGosset2016} and the SWAP method to compute the $2$nd R\'{e}nyi entanglement entropy. }
\end{figure*}

\subsection{The Jensen gap}
Since the mean purity is pinned, the entire $m$-dependence of
$\langle S_2\rangle$ is the gap between the mean of the logarithm and the
logarithm of the mean---the Jensen gap,
\begin{align}
\langle S_2\rangle
& = -\ln\bigl\langle \mathrm{Tr}\rho_A^2 \bigr\rangle + J, \\
 J & = \ln\bigl\langle \mathrm{Tr}\rho_A^2\bigr\rangle
  - \bigl\langle \ln \mathrm{Tr}\rho_A^2\bigr\rangle \;\geq\; 0,
\end{align}
with the first term fixed at the Haar value; $J$ is sourced entirely by the
fluctuations of the purity across the ensemble.  For $m \geq 2$ these
fluctuations are small, so write the purity as
$\mathrm{Tr}\rho_A^2 = \langle\mathrm{Tr}\rho_A^2\rangle\,(1+\epsilon)$ with
$\epsilon = (\mathrm{Tr}\rho_A^2 - \langle\mathrm{Tr}\rho_A^2\rangle)/
\langle\mathrm{Tr}\rho_A^2\rangle$, so that $\langle\epsilon\rangle = 0$ and
$\langle\epsilon^2\rangle = \mathrm{Var}(\mathrm{Tr}\rho_A^2)/
\langle\mathrm{Tr}\rho_A^2\rangle^2$.  Expanding the logarithm to second
order, $\ln(1+\epsilon) = \epsilon - \tfrac12\epsilon^2 + \cdots$, and
averaging,
\begin{align}
\bigl\langle\ln\mathrm{Tr}\rho_A^2\bigr\rangle
&= \ln\langle\mathrm{Tr}\rho_A^2\rangle
  + \langle\epsilon\rangle - \tfrac12\langle\epsilon^2\rangle + \cdots \\&
= \ln\langle\mathrm{Tr}\rho_A^2\rangle - \tfrac12\langle\epsilon^2\rangle + \cdots,
\end{align}
the linear term dropping out.  Inserting this into the gap leaves the
second-order piece,
\begin{equation}
J \approx \tfrac12\langle\epsilon^2\rangle
        = \frac{\mathrm{Var}(\mathrm{Tr}\rho_A^2)}
               {2\langle\mathrm{Tr}\rho_A^2\rangle^2},
\end{equation}
so the gap is controlled by the purity variance.

\subsection{The $1/m^3$ law}
Which fluctuation survives is clearest in the purity-tensor form of
Eq.~\eqref{eq:purity_tensor}: for the equal-weight, random-phase
superposition the purity carries an overall $1/m^2$ from the
normalisation,
\begin{equation}
\mathrm{Tr}\rho_A^2 = \frac{1}{m^2}\sum_{ijk\ell}
e^{\,i(\phi_i-\phi_j+\phi_k-\phi_\ell)}\,\mathcal{T}_{ijk\ell},
\end{equation}
and averaging over the random phases retains only the index patterns whose
phases cancel: the \emph{direct} pairing $(i{=}j,\,k{=}\ell)$, giving
$\mathcal{T}_{iikk}=\mathrm{Tr}(\rho_A^{(i)}\rho_A^{(k)})$, and the
\emph{swap} pairing $(i{=}\ell,\,k{=}j)$, giving
$\mathcal{T}_{ikki}=\|\mathrm{Tr}_B\ket{\psi_i}\!\bra{\psi_k}\|_{\mathrm{HS}}^2$
(less the $i{=}j{=}k{=}\ell$ overcount).  Their \emph{off}-diagonal
($i\neq k$) parts are overlaps of \emph{distinct} components---the
cross-overlap and the $B$-side Hilbert--Schmidt
norm, both equal to $2^{-|A|}$ for stabilizers (here $c_\perp = 0$, so,
unlike the sub-maximal case of Eq.~\eqref{eq:orthogonality}, the
swap term is full strength and supplies the second half of the pinned Haar
mean).  These self-average, their fluctuations vanishing exponentially in
$|A|$.  The rank deficit $\delta$, an $O(1)$ integer that does \emph{not}
shrink with subsystem size, survives only in the \emph{diagonal} ($i=k$)
self-overlap
$\mathcal{T}_{iiii}=\mathrm{Tr}(\rho_A^{(i)\,2})=2^{-(|A|-\delta_i)}$.  The
fluctuating part of the purity is therefore just this diagonal piece,
\begin{equation}
\bigl(\mathrm{Tr}\rho_A^2\bigr)_{\mathrm{fluct}}
= \frac{1}{m^2}\sum_{i=1}^{m}\mathcal{T}_{iiii}
= \frac{2^{-|A|}}{m^2}\sum_{i=1}^{m} 2^{\delta_i},
\end{equation}
a sum of $m$ \emph{independent} deficits $\delta_i$.  Pulling the constant
prefactor out of the variance and using independence to add the per-term
variances,
\begin{align}
\mathrm{Var}(\mathrm{Tr}\rho_A^2)
&= \Bigl(\frac{2^{-|A|}}{m^2}\Bigr)^{\!2}
   \mathrm{Var}\!\Bigl(\sum_{i=1}^{m} 2^{\delta_i}\Bigr)
 = \frac{2^{-2|A|}}{m^4}\sum_{i=1}^{m}\mathrm{Var}\bigl(2^{\delta_i}\bigr)
   \notag\\[2pt]
&= \frac{2^{-2|A|}}{m^4}\cdot m\,\mathrm{Var}(2^\delta)
 = 2^{-2|A|}\,\frac{\mathrm{Var}(2^\delta)}{m^3},
\end{align}
the $m$ identical deficit variances summing to $m\,\mathrm{Var}(2^\delta)$.
The $1/m^3$ is thus $(1/m^2)^2 \times m$: the square of the self-term's
$1/m^2$ weight times the $m$ independent components in the sum.  This is the
key identification---the entire gap is fixed by the single number
$\mathrm{Var}(2^\delta)$, the only ensemble quantity we now need to
compute.

\subsection{The deficit variance}
To assemble the moments we must be careful about what is random.  Fix one
realisation---a single random stabilizer state, with group $\mathcal{G}$.
Recall that the rank deficit $\delta$ is the number of
independent stabilizer generators supported fully on $A$.  These $\delta$
generators produce a subgroup of $2^\delta$ Paulis, all supported on $A$,
so $2^\delta$ is simply the number of $A$-only Paulis contained in
$\mathcal{G}$.  Counting them one Pauli at a time,
\begin{equation}
2^\delta = \!\!\sum_{P\,\in\,A\text{-Paulis}}\!\! \mathbf{1}[P\in\mathcal{G}],
\qquad
\mathbf{1}[P\in\mathcal{G}] = \begin{cases}1 & P\in\mathcal{G}\\ 0 & \text{else,}\end{cases}
\end{equation}
the sum running over all $4^{|A|} = d^2$ Paulis supported on $A$ (the
identity and $d^2-1$ nontrivial ones).  This count varies from realisation
to realisation, and the quantity we are after is its \emph{ensemble
variance}
\begin{equation}
\mathrm{Var}(2^\delta)
= \langle 4^\delta\rangle - \langle 2^\delta\rangle^2,
\qquad 4^\delta = \bigl(2^\delta\bigr)^2,
\end{equation}
so we need the first two ensemble averages.  Averaging the indicator form
over random $\mathcal{G}$ replaces each indicator by its probability, and, at
the balanced cut $2^N = d^2$,  Eqs.~\eqref{eq:prob_single} and
\eqref{eq:prob_pair} read $\Pr(P\in\mathcal{G}) = 1/(d^2+1)$ and
$\Pr(P,Q\in\mathcal{G}) = 1/[(d^2+1)(d^2/2+1)]$.

The mean is the single sum
\begin{align}
\langle 2^\delta\rangle
& = \!\!\sum_{P\,\in\,A\text{-Paulis}}\!\! \Pr(P\in\mathcal{G})  \notag \\  &
= \underbrace{1}_{\text{identity}} + (d^2-1)\,\frac{1}{d^2+1}
= \frac{2d^2}{d^2+1} \;\xrightarrow{}\; 2,
\end{align}
the identity always present and each of the $d^2-1$ nontrivial $A$-Paulis
contributing $1/(d^2+1)$.  The second moment squares that count.  Since
\begin{equation}
4^\delta = \bigl(2^\delta\bigr)^2
= \Bigl(\sum_P \mathbf{1}[P\in\mathcal{G}]\Bigr)^{\!2}
= \sum_{P,Q}\mathbf{1}[P\in\mathcal{G}]\,\mathbf{1}[Q\in\mathcal{G}]
\end{equation}
runs over all \emph{ordered} pairs $(P,Q)$ of $A$-Paulis---its summand equal
to $1$ only when \emph{both} lie in $\mathcal{G}$---averaging gives
$\langle 4^\delta\rangle = \sum_{P,Q}\Pr(P,Q\in\mathcal{G})$.  This splits by
case.  If both are trivial the pair is certain ($1$).  Three families
reduce to a \emph{single} nontrivial membership, and so carry the same
weight $1/(d^2+1)$ as the mean: $P$ trivial with $Q$ nontrivial, $Q$
trivial with $P$ nontrivial, and $P=Q$ nontrivial---each a family of
$d^2-1$ ordered pairs, giving $3(d^2-1)/(d^2+1)$ in total.  Finally, the
genuinely distinct nontrivial pairs ($P\neq Q$, both nontrivial) contribute
only when they commute---each nontrivial $A$-Pauli commutes with $d^2/2$ of
the $d^2$, i.e.\ $d^2/2-2$ nontrivial partners besides itself---each weighted by the pair probability $1/[(d^2+1)(d^2/2+1)]$.  Inserting these,
\begin{align}
\langle 4^\delta\rangle
&= \underbrace{1}_{\text{both triv.}}
 + \underbrace{3(d^2-1)\cdot\frac{1}{d^2+1}}_{\text{one triv. or }P=Q}
 \\ & + \underbrace{(d^2-1)\Bigl(\tfrac{d^2}{2}-2\Bigr)\cdot
   \frac{1}{(d^2+1)(\tfrac{d^2}{2}+1)}}_{\text{commuting distinct}}
   \notag\\
&= 1 + \frac{3(d^2-1)}{d^2+1}
 + \frac{(d^2-1)(d^2-4)}{(d^2+1)(d^2+2)}
 \\& = 1 + \frac{2(d^2-1)(2d^2+1)}{(d^2+1)(d^2+2)}
 \;\xrightarrow{}\; 5,
\end{align}
where the last term is (number of commuting distinct pairs) $\times$ (pair
probability): multiplying numerator and denominator by $2$ turns
$\tfrac{d^2}{2}-2$ into $d^2-4$ and the probability's $\tfrac{d^2}{2}+1$ into
$d^2+2$, which is where that denominator factor comes from.  Hence
$\mathrm{Var}(2^\delta) = 5 - 2^2 \to 1$ (and the mean
$\langle 2^\delta\rangle \to 2$ is the value used in the pinning check
above).  Dividing the
variance by $\langle\mathrm{Tr}\rho_A^2\rangle^2 = 4\cdot 2^{-2|A|}$,
in which the size factors cancel,
\begin{equation}
J(m) \approx \frac{\mathrm{Var}(2^\delta)}{8\,m^{3}}
\;\to\; \frac{1}{8\,m^{3}},
\end{equation}
in nats.  At the opposite extreme $m = 1$ the purity is $2^{-(|A|-\delta)}$
and the gap is exact,
\begin{align}
J(1) &= \ln\langle 2^\delta\rangle - \langle\delta\rangle\ln 2
\;\to\; (1 - \langle\delta\rangle)\ln 2 \notag \\ & \approx 0.10
\;(\approx 0.14~\text{bits}).
\end{align}
Collecting the two pieces, the main-text gain [Eq.~\eqref{eq:main_result}] is $\Delta S(m) = J(m)-J(1)$, the pinned Haar mean cancelling in the difference; it falls from $0$ at $m=1$ to $-J(1)\approx-0.10$ as $m\to\infty$.
The relaxation is essentially complete by $m = 2$: the
generic-component case is a pure fluctuation effect, in contrast to the
genuine spectral collapse of Appendix~\ref{app:filtered}.

\bibliography{main}

\end{document}